\documentclass{PoS}

\usepackage{epsfig}
\newcommand{\beqa}{\begin{eqnarray}}
\newcommand{\eeqa}{\end{eqnarray}}
\newcommand{\beq}{\begin{equation}}
\newcommand{\eeq}{\end{equation}}

\title{Large volume behaviour of Yang-Mills propagators}
\ShortTitle{Large volume behaviour of Yang-Mills propagators}

\author{\speaker{Christian S. Fischer}%
	 \\
        Institut f\"ur Physik, Technische Universit\"at Darmstadt,
	Schlossgartenstr. 9,\\
	64289 Darmstadt, Germany\\
        E-mail: \email{christian.fischer@physik.tu-darmstadt.de}}
\author{Reinhard Alkofer\\
Institut f\"ur Physik,
Karl-Franzens-Universit\"at, 
Universit\"atsplatz 5,
A-8010 Graz, Austria}
\author{Axel Maas\\
Institute of Physics,
Slovak Academy of Sciences,
D\'{u}bravsk\'{a} cesta 9,\\
SK-845 11 Bratislava, Slovakia}
\author{Jan M. Pawlowski\\
Institut f\"ur Theoretische Physik,
 University of Heidelberg,
  Philosophenweg 16,\\
  62910 Heidelberg, Germany}
\author{Lorenz von Smekal\\
Centre for the Subatomic Structure of Matter,
School of Chemistry and Physics,\\
The University of Adelaide,
SA 5005, Australia}

\abstract{We summarise results on finite-volume effects in the 
  propagators of Landau gauge Yang-Mills theory using Dyson-Schwinger 
  equations on a 4-dimensional torus. We demonstrate explicitly how
  the solutions for the gluon and the ghost propagator tend towards
  their respective infinite volume forms in the corresponding limit.
  We discuss the relation of our solutions with results
  from lattice Monte-Carlo simulations.}

\FullConference{The XXV International Symposium on Lattice Field Theory\\
		 July 30-4 August 2007\\
		 Regensburg, Germany}

\begin{document}

\section{Introduction}
  In Landau gauge continuum Yang-Mills theory there are two
  confinement scenarios connected to the infrared behaviour 
  of the gluon and ghost propagators: the Kugo-Ojima scenario 
  \cite{Kugo1979} and the Gribov-Zwanziger picture \cite{Gribov:1977wm}.
  Both predict an infrared enhanced ghost propagator and an infrared 
  suppressed gluon propagator (for reviews see \cite{Alkofer:2000wg}).
  In the Kugo-Ojima scenario BRST-symmetry is used to define
  a physical subspace of BRST-singlets within the complete state 
  space of covariant gauge QCD. Provided global gauge symmetry
  is unbroken one can show that the space of BRST-singlets indeed
  contains colourless objects only. This condition is connected 
  to the infrared behaviour of the ghost dressing function in Landau 
  gauge QCD: The global colour charge is well-defined if and only if the ghost 
  dressing function is singular in the infrared.
  This so called Kugo-Ojima condition is necessary, but not 
  sufficient, for confinement in the Kugo-Ojima scenario.
  The Gribov-Zwanziger scenario, on the other hand, postulates 
  that gauge field configurations on the Gribov-horizon are
  responsible for the confining nature of the quark-antiquark 
  potential. In terms of Green's functions these gauge field 
  configurations have been identified to induce infrared 
  enhancement in the ghost and infrared suppression of the gluon 
  propagator \cite{Gribov:1977wm,Zwanziger:2001kw,Gattnar:2004bf}.
  
  Studies of (untruncated) Dyson-Schwinger equations (DSE) and functional 
  renormalisation group equations (FRGE) in the infinite volume/continuum 
  limit strongly support these scenarios.
  In Landau gauge the ghost and gluon propagators are given by
 \beq
 D^G(p^2) = - \frac{G(p^2)}{p^2} \, , \qquad
 D_{\mu \nu}(p^2)  = \left(\delta_{\mu \nu} -\frac{p_\mu 
 p_\nu}{p^2}\right) \frac{Z(p^2)}{p^2} \, .
 \eeq
 where $G(p^2)$ denotes the ghost dressing function and $Z(p^2)$
 the dressing function of the gluon. One finds \cite{vonSmekal1997} 
 that the small momentum 
 behaviour of these functions is given by power laws, i.e.
 \beq
 G(p^2) \sim (p^2)^{-\kappa}, \hspace*{1cm} Z(p^2) \sim (p^2)^{2\kappa}\,.
 \label{kappa}
 \eeq Here the exponents of ghost and glue are uniquely related by the
 anomalous dimension $\kappa$. Similar expressions have been found for
 all other one-particle irreducible (1PI) Greens's functions in Landau
 gauge \cite{Alkofer:2004it,Fischer:2006vf}; see section \ref{1PI}.
 In the notation (\ref{kappa}) the Kugo-Ojima and Gribov-Zwanziger
 scenarios translate to the condition $\kappa \ge 1/2$. Indeed, in the
 DSE and FRGE approaches one obtains $\kappa = (93 - \sqrt{1201})/98
 \approx 0.595$
 \cite{Zwanziger:2001kw,Lerche:2002ep,Pawlowski:2003hq}, which
 satisfies both criteria.

 On the lattice the verification of the relations (\ref{kappa}) in
 gauge fixed calculations turned out to be extremely cumbersome, for
 latest results see
 \cite{Cucchieri:2007ta,Cucchieri:2006xi,Bogolubsky:2007bw,Bowman:2007du,Bowman:2004jm,Oliveira:2007dy,Sternbeck:2005tk,Sternbeck:2006rd,Oliveira:2006yw,Sternbeck:2006cg}.
 To extract information on the power laws (\ref{kappa}) from studies
 on a finite volume it is most important to address the volume
 dependence of the long-range behaviour of these correlations.  It is
 only when this dependence is under control that firm conclusions can
 be drawn from extrapolations to the infinite volume and continuum
 limits.

 In order to understand possible patterns of such volume effects it is 
 an obvious and necessary step forward to adapt the continuum methods 
 to finite volumes \cite{Fischer,Fischer:2007pf}. We summarise these 
 efforts in section \ref{torus} and compare the results to available 
 lattice data.

\section{Infrared exponents of 1PI Green's functions \label{1PI}}

The infrared behaviour of the one-particle irreducible (1PI) Green's
functions of Yang-Mills theory have been investigated in a number of
works. The basic relation (\ref{kappa}) between the dressing functions
of the gluon and ghost propagator has been extracted first in
\cite{vonSmekal1997}. These findings have been generalised to Green's
functions with an arbitrary number of legs in \cite{Alkofer:2004it}.
The analysis rests upon a separation of scales, which takes place in
the deep infrared momentum region. Provided there is only one external
momentum $p^2$ much smaller than $\Lambda_{\mathrm{QCD}}$, a
self-consistent infrared asymptotic solution of the whole tower of
DSEs or FRGEs for these functions is given by \beq \Gamma^{n,m}(p^2)
\sim (p^2)^{(n-m)\kappa}. \label{IRsolution} \eeq Here
$\Gamma^{n,m}(p^2)$ denotes the dressing function of the infrared
leading tensor structure of the 1PI-Green's function with $2n$
external ghost legs and $m$ external gluon legs. It is important to
note that the solution (\ref{IRsolution}) is unique
\cite{Fischer:2006vf} which follows from a comparison of untruncated
towers of DSEs and FRGEs. The exponent $\kappa$ is known to be
positive \cite{Watson:2001yv,Lerche:2002ep}.  We emphasise that these
findings do not rely on any truncation scheme.

This specific value of $\kappa = (93 - \sqrt{1201})/98 \approx 0.595$,
however, depends on the assumption that a bare ghost-gluon vertex is a good 
approximation to the full vertex in the infrared. This assumption has been 
tested in the continuum \cite{Schleifenbaum:2004id} and on the lattice 
\cite{Sternbeck:2006rd,Maas:2006qw}, and found to
be adequate. Possible corrections by regular dressings in the infrared have
been investigated within the DSE framework in \cite{Lerche:2002ep}, where a 
possible range $0.5 \le \kappa < 0.7$ has been given. Note that the value
$\kappa=1/2$ marks the watershed between an infrared vanishing ($\kappa >1/2$)
and an infrared divergent ($\kappa < 1/2$) gluon propagator. The first 
option necessarily entails that the gluon propagator violates positivity
as can be seen from
\beq
  0 = D(p=0) = \int {d^4x} \; \Delta(x) \,, \label{zero}
\eeq
with $D(p) = Z(p^2)/p^2$. This relation implies that the propagator
function in coordinate space, the Schwinger function $\Delta(x)$, must
contain positive as well as negative norm contributions, with equal
integrated strengths. Therefore infrared vanishing gluons cannot be
part of the positive definite, physical state space of Yang-Mills theory.
While the infrared vanishing of the gluon propagator is not at all
necessary for the positivity violations of transverse gluons, one
needs to analyse the convexity in Euclidean time of its
one-dimensional Fourier transform to establish this, if the gluon is
infrared finite or even divergent.

Further interesting consequences of the solution (\ref{IRsolution}) are the 
existence of infrared fixed points in the running couplings of Yang-Mills 
theory \cite{vonSmekal1997,Alkofer:2004it}. 

\section{Dyson-Schwinger equations on a torus \label{torus}}

There are several caveats in comparing results from the continuum
Dyson-Schwinger approach to those of lattice calculations. First, the 
quantitative aspects of the continuum solutions (i.e. the value of $\kappa$
and numerical results at intermediate momenta) depend on the details
of the chosen truncation scheme, whereas the lattice calculations are 
{\it ab initio}. 
On the other hand, simulations are performed on finite lattices and one 
has to deal with the effects due to the finiteness of volume and 
lattice spacing.  

To quantify the 'plain' volume effects (i.e. those not connected to
the gauge fixing procedure) we formulated the DSEs on a torus
\cite{Fischer,Fischer:2007pf}, employing the same truncation scheme as
in the infinite volume/continuum framework \cite{fa}. One would then
expect to see differences to the continuum solution for small volumes,
which disappear continuously when the volume is chosen larger and
larger. This expectation is supported by the FRGE infrared studies
with an explicit infrared cut-off \cite{Pawlowski:2003hq}. In earlier
works on DSEs on the torus \cite{Fischer} this issue could not be
resolved but an improved finite-volume renormalisation procedure indeed 
led to a smooth transition \cite{Fischer:2007pf}. 

\begin{figure}[th!]
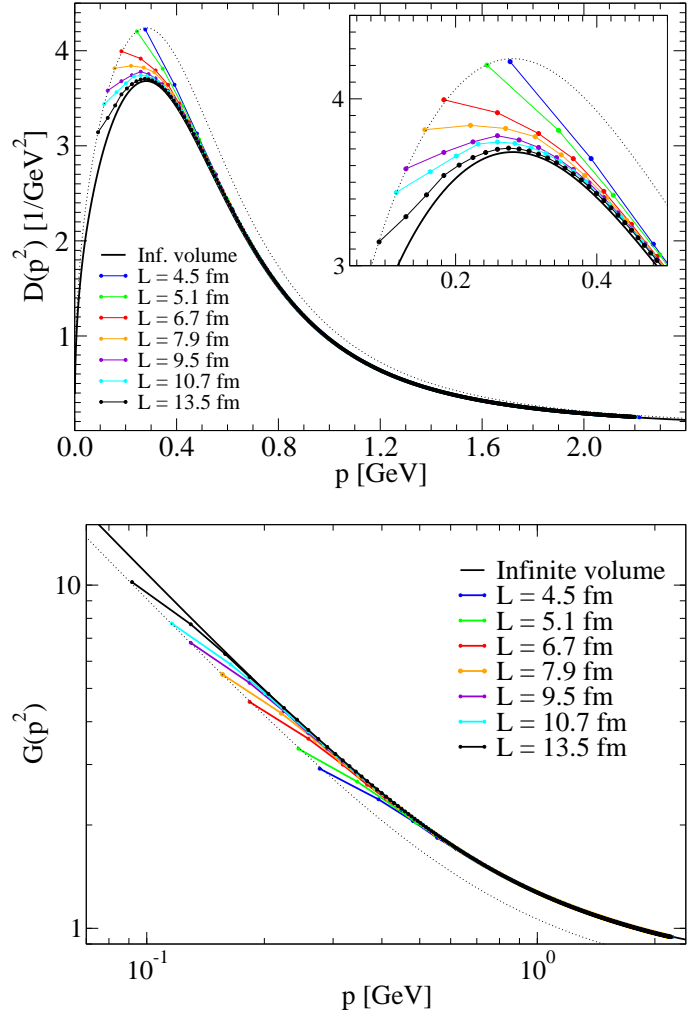

\centerline{
\epsfig{file=Fig3a.eps,width=9cm}}
\vspace*{4mm}
\centerline{
\epsfig{file=Fig3b.eps,width=9cm}}
\caption{Numerical solutions on tori with different volumes compared
  to the infinite volume limit. The upper graph shows the
  gluon propagator, whereas on the lower graph the ghost dressing
  function is plotted.}\label{Fig2}
\end{figure}

The corresponding numerical results on different volumes are shown in 
Figure \ref{Fig2}. The momentum scale is fixed by comparison with 
corresponding lattice calculations, see \cite{Fischer:2007pf} for 
details. We discuss results on seven different volumes $V=L^4$; the
corresponding box lengths $L$ are given in the legends of Figure
\ref{Fig2}. One clearly observes that the infinite volume solutions of
the gluon propagator $D(p^2)$ and ghost dressing function $G(p^2)$ are
more and more approached by the torus solutions with increasing
volume. Qualitatively one can see the following behaviour: the
gluon propagator seems to be divergent at volumes of $V \approx (4-8
\,\mbox{fm})^4$. For larger volumes the propagator bends downwards 
to reach a plateau at roughly $V \approx (9 \,\mbox{fm})^4$ and at $V
\approx (10 \,\mbox{fm})^4$ the propagator is infrared vanishing and
therefore qualitatively similar to the infinite volume limit given in
(\ref{kappa}). For the ghost we observe that the first two points on 
each volume bend away from the power law behaviour of the infinite 
volume solution. With increasing volume more and more of the remaining 
points are in the 'scaling region', where the infinite volume power 
law develops. 

\begin{figure}[t]
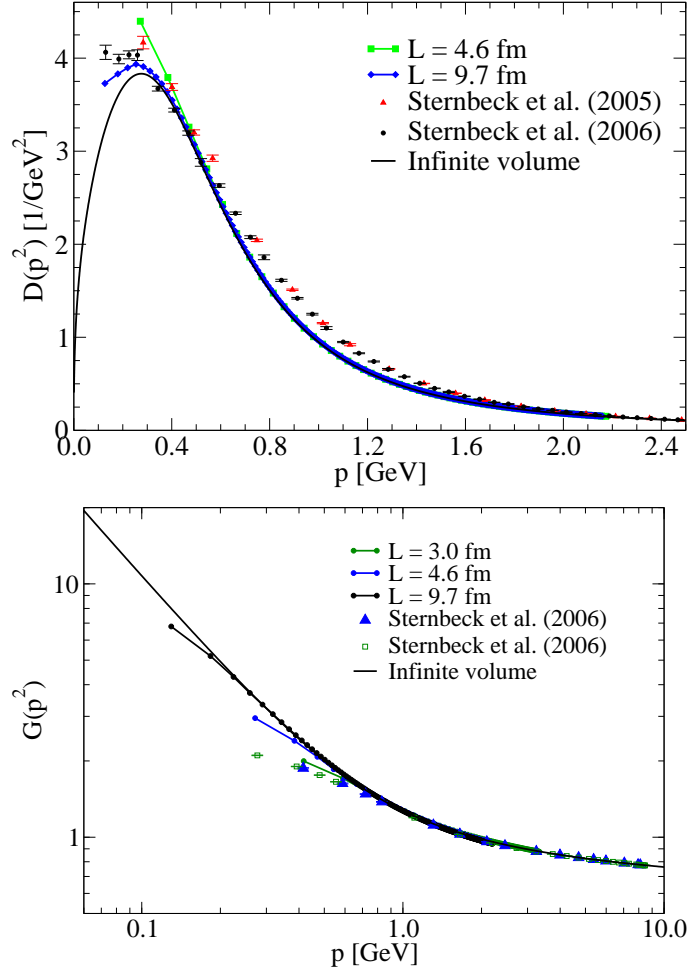

  \centerline{ \epsfig{file=Fig8a.eps,width=9cm}} \vspace*{2mm}
  \centerline{ \epsfig{file=Fig8b.eps,width=9cm}}
\caption{DSE results for the gluon propagator and ghost dressing
  function on tori with different volumes compared to recent lattice
  calculations on similar manifolds. The lattice data are taken from
  Refs.~\cite{Sternbeck:2006cg,Sternbeck:2005tk}.
\label{fig:latt}}    
\end{figure}

We now compare our results from DSEs on a torus to the ones from
lattice calculations. Nowadays, lattice data for the
gluon propagator are available on impressively large lattices. The
authors of \cite{Cucchieri:2006xi} report on an $SU(2)$-study on a
$52^4$-lattice, whereas in \cite{Sternbeck:2006cg} results from an
$SU(3)$ calculation on a $56^4$-lattice are discussed. Results
on even larger lattices have been presented at this conference 
\cite{Cucchieri,Sternbeck}. In the upper graph of 
Figure~\ref{fig:latt} we display the $SU(3)$-results of 
\cite{Sternbeck:2006cg} together with data on a smaller volume
\cite{Sternbeck:2005tk} and compare with DSE-results on tori with
similar volumes.

The qualitative agreement of the solutions at similar volumes in the
infrared is interesting.\footnote{Differences in the intermediate
  momentum regime at approximately 1 GeV are truncation artefacts of the DSE
  solutions. One can show analytically that this is the only region
  where the omitted gluonic two-loop diagrams contribute
  significantly.}  Whereas both, the lattice and the DSE result at the
smaller volume $V \approx (4.6 \,\mbox{fm})^4$ seem to diverge, one
starts to observe an infrared finite one, or perhaps even a slight
infrared suppression, at the larger volumes $V \approx (9.7
\,\mbox{fm})^4$.  This indicates that the scaling behaviour of the
lattice results with volume may be similar to the ones of the DSE
solution. If this is correct one should see a turnover of the gluon 
propagator at even larger volumes.  

The situation is far less clear for the ghost dressing function. 
Our results for three different volumes, $V = (3.9, 4.6, 9.7 \,{\rm fm})^4$, 
are compared to the SU(3) lattice results of
\cite{Sternbeck:2006cg,Sternbeck:2005tk}. For the DSE solutions we 
observe a characteristic deviation of the two lowest momentum points 
at each volume from the infinite volume solution, corresponding to a 
ghost mass which goes to zero in the infinite volume limit \cite{Fischer:2007pf}. 
The lattice results do not seem to show such behaviour as yet. Even 
though the lattice volumes herein are roughly between 3 fm and 4.5 fm, 
and thus still rather small, there appears to be not much sign of a 
volume dependence at all at this point. However, it has been observed
that effects at intermediate momenta can in a subtle way influence the
finite-volume effects in the ghost propagator \cite{Maas:2007uv}. As these
are truncation-dependent in the DSEs, this may explain the difference to
the results in lattice calculations.

\section{Epilogue}

In this conference first results have been presented for the gluon
propagator on very large lattices ($128^4$ \cite{Cucchieri} and
$112^4$ \cite{Sternbeck}).  Both groups do not see a turnover of the
gluon propagator, even though the volumes exceed $(10 \,\mbox{fm})^4$ by
far. Does this mean that the gluon propagator is finite in the
infrared, i.e. $\kappa=0.5$ ? In principle there is not much to say
against this scenario from the pure DSE/FRGE-perspective. It could be
that the IR-regular dressing of the nonperturbative ghost-gluon vertex
provides the necessary corrections to drive the IR-anomalous dimension
towards $\kappa \cong 0.5$. However, it is an unambiguous and
truncation independent prediction of the Dyson-Schwinger and the
functional renormalisation group framework that the ghost dressing
function should then diverge as $G(p^2) \sim (p^2)^{-0.5}$
\cite{Fischer:2006vf}. Current lattice results in four dimensions do
not seem to support this relation. On the other hand, lattice data in
two dimensions \cite{Maas:2007uv} agree well with the corresponding
DSE-results \cite{Huber:2007kc}.
The usual suspects for the remaining discrepancy in four dimensions
then being gauge fixing or renormalisation problems.  Gauge fixing
algorithms are known to be less efficient at larger lattices where the
number of Gribov copies increases exponentially. Naturally this effect
also depends on the number of dimensions.  Furthermore, it is also
known, that effects from Gribov copies influence the ghost
propagator much stronger than the glue
\cite{Sternbeck:2005tk,Oliveira:2006yw}. Finally, very recent lattice
results
allowing a wider class of gauge transformations in the minimisation
procedure \cite{Bogolubsky:2007bw} tend to result in an additional
suppression of the gluon propagator at low momenta, which is observed
already in rather moderate volumes of $V=(6.5 \,\mbox{fm})^4$.

Also in the continuum one probably needs additional verification for
Zwanziger's idea that gauge fixing on the first Gribov region (as done
in DSEs and FRGEs) is enough to avoid effects from Gribov copies
\cite{Zwanziger:2001kw}. This idea underlies all continuum studies of
infrared anomalous dimensions so far. Therefore, the final settlement
of the infrared behaviour of ghost and glue awaits further
clarification in both, the continuum and lattice studies.

\noindent{\bf Acknowledgements}

CF would like to thank the organisers of {\it Lattice 2007\/}
for all their efforts which made this inspiring conference possible.
We are grateful to A.~Cucchieri, T.~Mendez, 
M.~M\"uller-Preussker, O.~Oliveira, A.~Sternbeck and A.~Williams 
for interesting discussions. 
This work was supported by the 
Helmholtz-University Young Investigator Grant VH-NG-332, by the DFG
under grant no.\ MA-3935/1-2 and Al 279/5-2, 
and by the Australian Research Council.

\end{document}